%% file: paper.tex
\documentclass[sigplan,10pt,screen,nonacm]{acmart}
\renewcommand\footnotetextcopyrightpermission[1]{}
\usepackage{xspace}
\usepackage{tabularx}
\usepackage{subcaption}
\usepackage{enumitem}
\usepackage{algorithm}
\usepackage{algorithmicx}
\usepackage[noend]{algpseudocode}
\usepackage{pifont}
\usepackage{multirow}
\usepackage{booktabs}
\usepackage{graphicx}
\usepackage{etoolbox}
\usepackage[all]{nowidow}
\usepackage{colortbl}
\usepackage{latexsym}
\usepackage{svg}
\usepackage{rotating}
\usepackage{amsmath}

\setitemize{noitemsep,topsep=0pt,parsep=0pt,partopsep=0pt}

\newcommand{\parabf}[1]{\medskip\noindent\textbf{#1}}

\newcommand{\cut}[1]{}

\newcommand{\sysname}{TokenLake\xspace}

\begin{document}
\title{\sysname: A Unified Segment-level Prefix Cache Pool for Fine-grained Elastic Long-Context LLM Serving}
\pagestyle{plain}
\author{
\textit{Bingyang Wu$^{1}$\qquad Zili Zhang$^{1}$\qquad Yinmin Zhong$^{1}$\qquad Guanzhe Huang$^{2}$\\\qquad Yibo Zhu$^{2}$\qquad
Xuanzhe Liu$^{1}$\qquad Xin Jin$^{1}$\\
$^{1}$\textit{Peking \  University} \qquad $^{2}$\textit{StepFun}}
}

\input{sections/abstract}

\maketitle

\input{sections/introduction}
\input{sections/background}

\input{sections/overview}
\input{sections/design}

\input{sections/evaluation}
\input{sections/related}
\input{sections/conclusion}

\bibliographystyle{style/ACM-Reference-Format}
\bibliography{paper}

\end{document}

%% file: sections/abstract.tex
\begin{abstract}
Prefix caching is crucial to accelerate multi-turn interactions and requests with 
shared prefixes.
At the cluster level, existing prefix caching systems are tightly coupled with request scheduling to optimize cache efficiency and computation performance together, leading to load imbalance, data redundancy, and memory fragmentation of caching systems across instances. To address these issues, memory pooling is promising to shield the scheduler from the underlying cache management so that it can focus on the computation optimization.
However, because
existing prefix caching systems
only transfer increasingly longer prefix caches between instances, they cannot achieve low-latency memory pooling.

To address these problems, we propose a unified segment-level prefix cache pool, \sysname. It uses a declarative cache interface to expose requests' query tensors, prefix caches, and cache-aware operations to \sysname for efficient pooling.
Powered by this abstraction, \sysname can manage prefix cache at the segment level with a heavy-hitter-aware load balancing algorithm to achieve better cache load balance, deduplication, and defragmentation. \sysname also transparently minimizes the communication volume of query tensors and new caches. Based on \sysname, the scheduler can schedule requests elastically by using existing techniques without considering prefix cache management.
Evaluations on real-world workloads show that \sysname can improve throughput by up to 2.6$\times$ and 2.0$\times$ and boost hit rate by 2.0$\times$ and 2.1$\times$, compared to state-of-the-art cache-aware routing and cache-centric PD-disaggregation solutions, respectively.

\end{abstract}

%% file: sections/introduction.tex
\section{Introduction}
\label{sec:introduction}

With the advancements of Large Language Models (LLMs), the context lengths of LLMs are becoming increasingly longer to support more sophisticated tasks~\cite{gemini, claude4, liu2023world}.
This trend is reflected in increasingly \textit{longer input prompts} and a growing number of \textit{multi-turn interactions} with users and external tools.
For example, 
agent application Manus reports their average input-to-output ratio is \textit{100:1}~\cite{manus}, and the average number of turns per session on
Character.AI is \textit{180 turns}~\cite{charactorai}.
To reduce the computation introduced by long context,
a key strategy is \textit{prefix caching}, specifically reusing previously computed KV cache, for requests with the shared prefix. 
Evidence from CharacterAI~\cite{charactorai} and Kimi~\cite{mooncake} reveals that the hit rate can reach up to 75\% to 95\% in production. As the KV cache size increases linearly with the sequence length, efficiently managing such a large volume of KV cache to achieve a high cache hit rate has become a critical problem.

A popular approach to maximize cache utilization is to use cache-aware routing~\cite{sglang,preble,aibrix,dynamo,vllm}. The prefix cache on each instance is organized as a local prefix tree. For each request, the router dispatches it to one instance based on the per-instance hit rates and load-balancing strategy.

However, it leads to the following issues.
(1) For instances with high hit rates, the memory bandwidth is over-utilized, whereas for instances with low hit rates, the memory bandwidth is under-utilized, leading to the \textit{load imbalance}.
(2) To balance the computation across instances, requests with shared prefixes, such as system prompts and shared documents, are often replicated across instances, leading to severe \textit{redundancy}.
(3) Because the entire prefix requires residing in the same instance to form a prefix tree, available slots across instances cannot be used together, leading to \textit{fragmentation}.

Meanwhile, to avoid interference between the prefill and decoding phase, PD disaggregation~\cite{zhong2024distserve, patel2024splitwise, hu2024inference} has been adopted to disaggregate these two phases on different instances. This further exacerbates the above problems.
Specifically, instances in different phases have different demands on memory bandwidth~\cite{flashinfer}, leading to the \textit{load imbalance}. In multi-turn scenarios, both prefill and decode instances must store the same prefix cache to avoid re-computation or repeated transfer, causing \textit{redundancy}. Similarly, the prefix cache slots on prefill and decode instances are hard to share by existing caching systems, leading to \textit{fragmentation}.

To fundamentally address these issues, GPU memory pooling is a promising solution. By aggregating all GPU memories into a unified pool and decoupling memory access from computation, it is possible to utilize all available GPU memory bandwidth to avoid load imbalance and all GPU memory as a single cache to eliminate redundancy and fragmentation, while supporting elastic scheduling to accommodate dynamic workloads without considering data locality.

However, existing prefix caching systems~\cite{mooncake,lmcache,aibrix,preble,memserve} are not efficient for GPU memory pooling.
Their interface, e.g., \texttt{put}, \texttt{get}, and \texttt{transfer}, requires the scheduler to consider cache management and can only imperatively transfer increasingly longer prefix cache across instances for computation. This architecture suffers from high communication overhead and binds two different and sometimes conflicting concerns together, i.e., cache management for high memory efficiency and scheduling for high computation efficiency of dynamic workloads, leading to suboptimal performance.

For the above problems, we propose a \textit{declarative} prefix cache interface and build a unified segment-level prefix cache pool, \sysname, to decouple prefix cache management and scheduling by exposing query tensors, prefix caches, and cache-aware operations to \sysname. This design allows \sysname to achieve low-latency memory pooling by transparently optimizing transmission of query tensors and prefix cache together at finer granularity, while enabling the scheduler to elastically schedule requests' cache-free operations, such as self-attention and feed-forward network module, in a stateless way for better performance.

However, efficiently utilizing all GPU resources is still non-trivial.
First, even if integrated query tensors and cache-aware operations into \sysname, a simple pooling strategy does not guarantee efficient utilization of all memory bandwidth. For example, if using a strict locality-based policy that spills the prefix cache to other instances only when the local memory is insufficient, it still leads to load imbalance and frequent migration of prefix cache (\S\ref{sec:load_balance}).
Second, finer-grained management dramatically expands the decision space of caching, making effective load balancing, duplication control, and communication optimization difficult.

To address these challenges, we introduce a heavy-hitter-aware segment-level load balancing algorithm to effectively utilize memory and bandwidth. We also propose a bipartite matching-based dispatching algorithm to minimize the communication volume of query tensors and newly generated KV cache under the given scheduler's decision.
Finally, we implement a goodput-optimized stateless elastic scheduling algorithm by using PD disaggregation, chunked prefill, and elastic sequence parallelism together to demonstrate its expressiveness and its potential to enable fine-grained scheduling.
The complexity of these algorithms is polynomial to the number of instances, making them efficient and scalable.

Evaluations show that, compared to state-of-the-art cache-aware routing, SGLang-Router~\cite{sglang} and cache-centric PD-disaggregation solutions, MoonCake~\cite{mooncake}, \sysname can improve throughput
under a given SLO by up to 2.6$\times$ and 2.0$\times$, and boost hit rate by up to 2.0$\times$ and 2.1$\times$, respectively.

In summary, our contributions are as follows:
\begin{itemize}[leftmargin=*]
    \item We propose a declarative prefix cache interface to decouple prefix cache management and request scheduling to achieve efficient memory pooling.
    \item We propose a heavy-hitter-aware segment-level load balancing algorithm and a bipartite matching-based dispatching algorithm to efficiently achieve load balance, lower redundancy, and less fragmentation of the prefix cache.
    \item We implement a goodput-optimized stateless elastic scheduling algorithm to demonstrate the expressiveness of the interface and conduct a comprehensive evaluation to show the effectiveness of \sysname.
\end{itemize}

%% file: sections/background.tex
\section{Background}
\label{sec:background}

\subsection{LLM serving systems}
The inference process of Large Language Models (LLMs) can be divided into two distinct phases: the prefill phase and the decoding phase. The prefill phase processes the entire input tokens in one iteration to generate the initial Key-Value (KV) cache, intermediate states for future iteration, while the decoding phase generates output tokens iteratively and stores newly generated KV cache in the prefix cache. For multi-turn sessions, they contain multiple turns of requests, each with its respective prefill and decoding phase. Especially, each turn also needs the KV cache of previous turns. In each iteration of the inference process, the prefix cache only involves the attention module. In each layer, query tensors generated from the input tokens use prefix attention and self-attention to interact with all preceding KV cache and KV cache of themselves, respectively, and then reduce the partial output tensors and normalizers to generate the final output tensors of the attention module. Therefore, except for the prefix attention, which is \textit{cache-aware}, self-attention and all other operations, such as feed-forward network (FFN) and projection layer, are \textit{cache-free}, i.e., computation only depends on the previous module's output and model parameters without directly using the prefix cache.

\begin{figure}[t]
    \includegraphics[width=\linewidth]{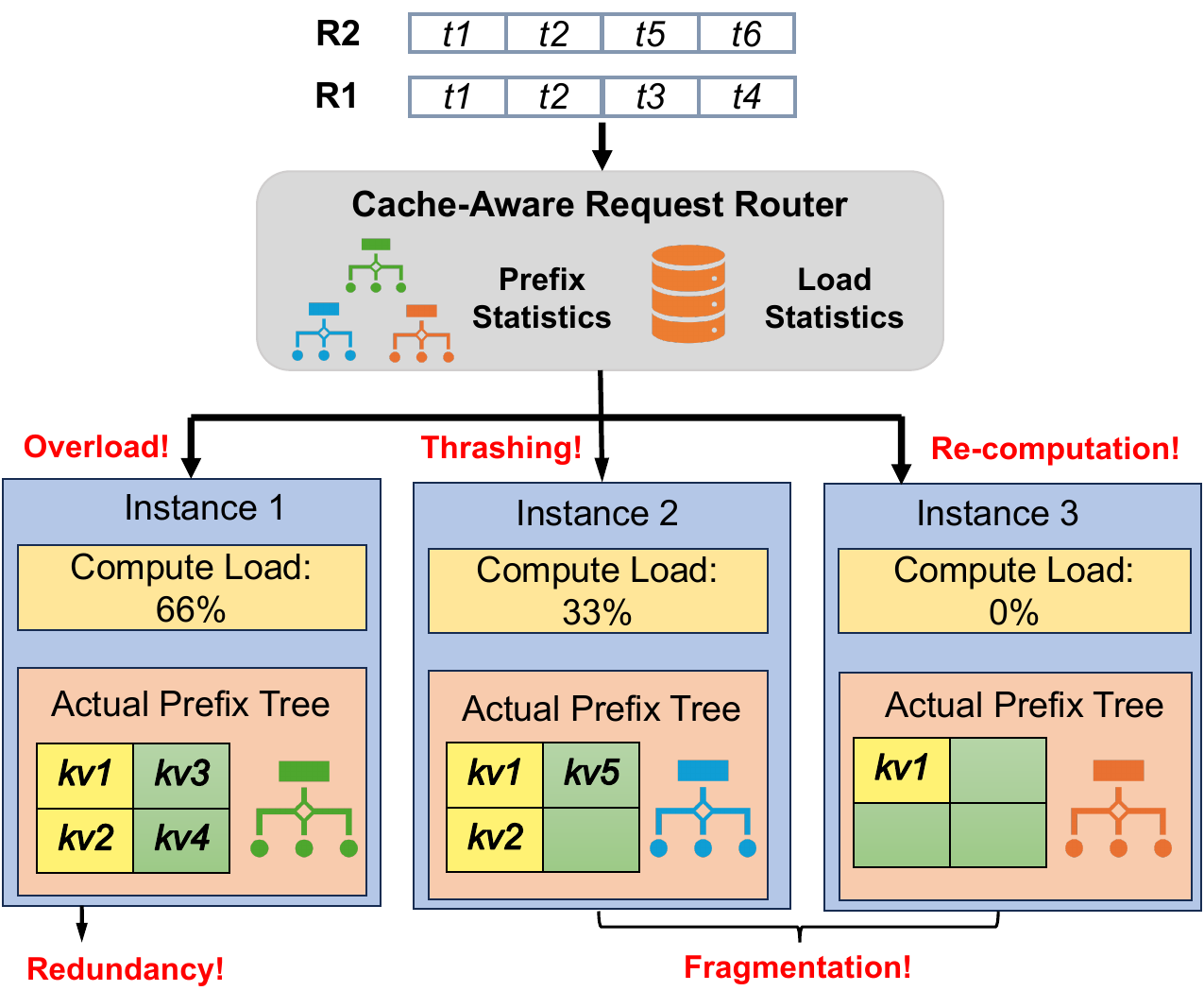}
    \caption{Limitations of cache-aware routing.}
    \label{fig:background:cache_aware_router}
\end{figure}

To support more applications, such as analyzing large codebases and stateful multi-turn agent interactions, the context length of LLMs has been significantly increased~\cite{gemini, claude4, liu2023world}.
To accelerate the input processing speed of long-context inference, prefix caching is proposed to further share KV caches across requests with the same prefix or multiple turns in a session~\cite{sglang,yu2025stateful, gao2024cost, pan2025kvflow, xu2024pie, zuo2025serving, jiang2024neo}. To quickly retrieve the relevant prefix cache, prefix caches are typically organized as a prefix tree in each LLM instance~\cite{sglang}.

At the same time, many scheduling algorithms have been proposed to accommodate the distinct characteristics of the prefill and decoding phases.
On the instance level, chunked prefill~\cite{agrawal2024taming, kamath2025pod, holmes2024deepspeed} is proposed to split the long context into smaller chunks and batch requests in two phases together to reduce the interference between them.
On the cluster level, PD disaggregation~\cite{wang2025prefill, qiao2024conserve, ruan2025dynaserve, jin2024system, du2025ecoserve, feng2025windserve, zhao2024blendserve} is proposed to disaggregate the prefill and decoding phases into different instances to eliminate interference between the two phases.
Elastic sequence parallelism (ESP)~\cite{wu2024loongserve} is further proposed to handle variable context lengths and frequent ratio changes between prefill and decoding phases by elastically setting the degree of parallelism (DoP) for requests with various context lengths in different phases.

\subsection{LLM caching systems}
To support prefix management on each instance and complex scheduling algorithms on the cluster level, LLM serving systems require efficient caching systems to manage the prefix cache. Existing caching systems typically expose an \emph{imperative} interface, such as \texttt{put}, \texttt{get}, and \texttt{transfer} APIs, for LLM serving systems to imperatively cache and retrieve a prefix cache in the local prefix tree and transfer it to another instance~\cite{mooncake,lmcache,aibrix,preble,memserve}. 

\begin{figure}[t]
    \includegraphics[width=\linewidth]{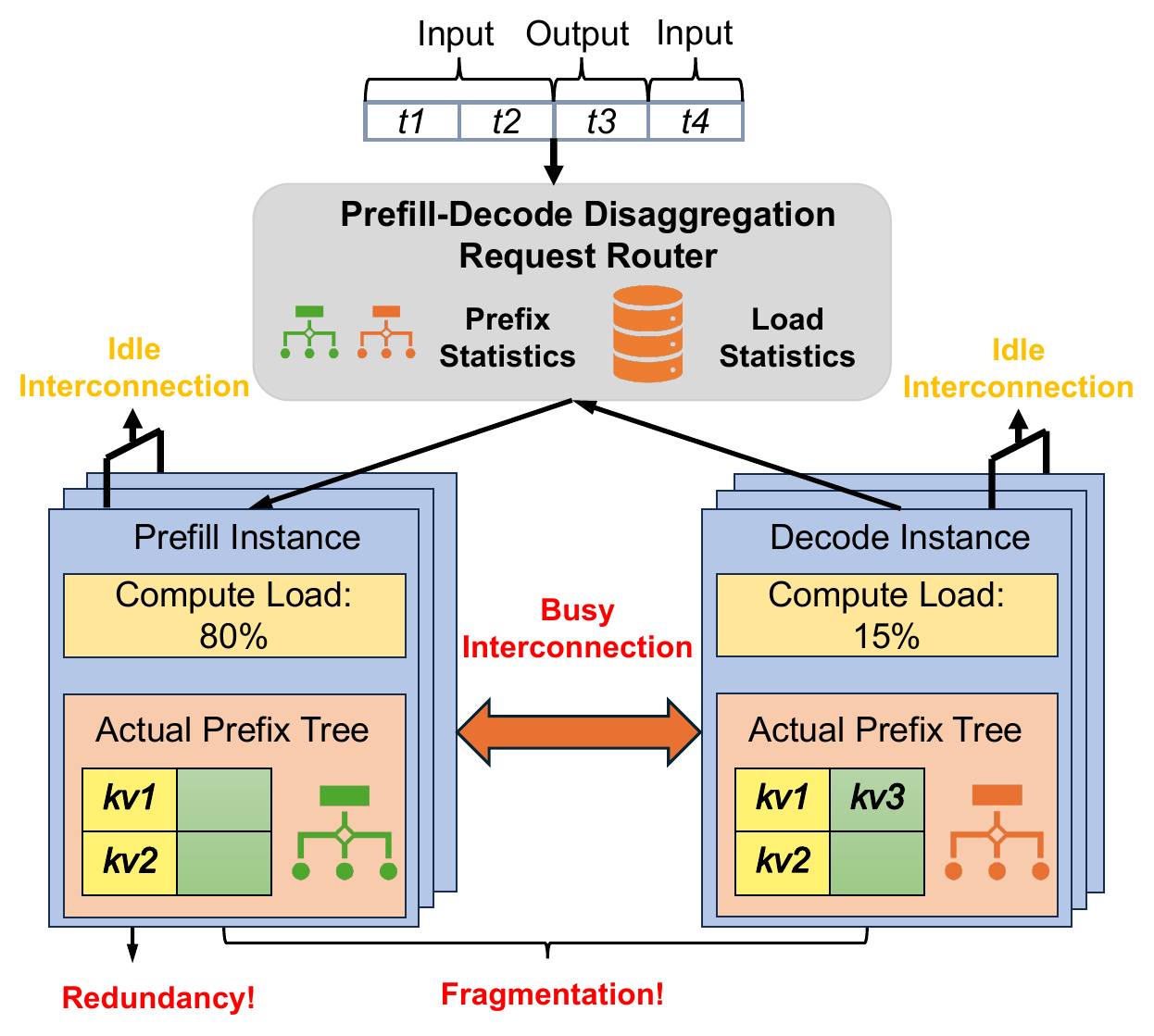}
    \caption{Limitations of cache-centric PD disaggregation.}
    \label{fig:background:pd_disaggregation}
\end{figure}

Because of the imperative nature, the caching system is tightly coupled with the scheduling algorithm.
The scheduler of the LLM serving system has to consider the state of the prefix cache in each instance for maximizing cache efficiency while improving requests' computation performance, such as throughput and latency.
The tight coupling between schedulers and their underlying imperative caching systems has given rise to two popular policies.

The first is cache-aware request routing. With the statistics collected from instances, the cache-aware router dispatches each incoming request to an instance by trading off between multiple potentially conflicting goals, such as maximizing cache hit rate and balancing computation across instances.

The other is cache-centric PD disaggregation. In this policy, the user decides the number of prefill and decode instances based on the computation and cache demands of the predicted workload. A PD disaggregation router is used to dispatch requests to prefill instances first, then imperatively invokes the \texttt{transfer} API to transfer the generated prefix cache to decode instances for subsequent decoding.

\subsection{Motivations}
\textbf{Limitations of existing solutions.}
To motivate our design, we first analyze the limitations of existing solutions.

A primary limitation is the severe \textit{load imbalance} inherent in the two mainstream scheduling policies.
For cache-aware routing, this tight coupling makes instances with high hit rate attract more requests computation, more KV cache generation and storage on the same instance, which further increases the hit rate of the instance, while other instances are under-utilized, leading to load imbalance in terms of number of accessed prefix cache, as shown in overloaded instance 1 of Figure~\ref{fig:background:cache_aware_router}.
For PD disaggregation, load imbalance is even more pronounced.
As shown in Figure~\ref{fig:background:pd_disaggregation}, in PD disaggregation, decode instances not only use prefix cache generated at the decoding phase, but also use prefix cache sent from the prefill phase, creating a fixed, asymmetric cache load that cannot be balanced between the two groups.

The second limitation is the \textit{data redundancy},
which leads to inefficient cache utilization. For cache-aware routing, load balancing policy forces the caching system to replicate shared prefix caches across multiple instances ($kv1$, $kv2$), and may introduce re-computation overhead to generate replicated prefix (e.g., dispatch $R1$ to instance 3 in Figure~\ref{fig:background:cache_aware_router}).
Similarly, PD disaggregation necessitates that the prefix cache generated by prefill instances is transferred to decode instances for subsequent computation, while caching them in prefill instances for incoming turns or requests, leading to redundancy ($kv1$, $kv2$).
In both scenarios, the redundancy is inevitable for existing imperative caching systems, otherwise it causes frequent transfer for increasingly longer prefix cache as the number of turns in sessions increases.

The last limitation is \textit{memory fragmentation}.
In the scenario depicted in Figure~\ref{fig:background:cache_aware_router} and Figure~\ref{fig:background:pd_disaggregation},
the system has four free slots in aggregate, but a new request (e.g., R3) requiring this amount cannot be placed because the cache capacity on any instance is insufficient. It leads to thrashing, i.e., eviction of prefix caches that may be required in the future ($kv5$ in instance 2 of Figure~\ref{fig:background:cache_aware_router}), introducing substantial overhead.

\begin{figure}[t]
    \includegraphics[width=\linewidth]{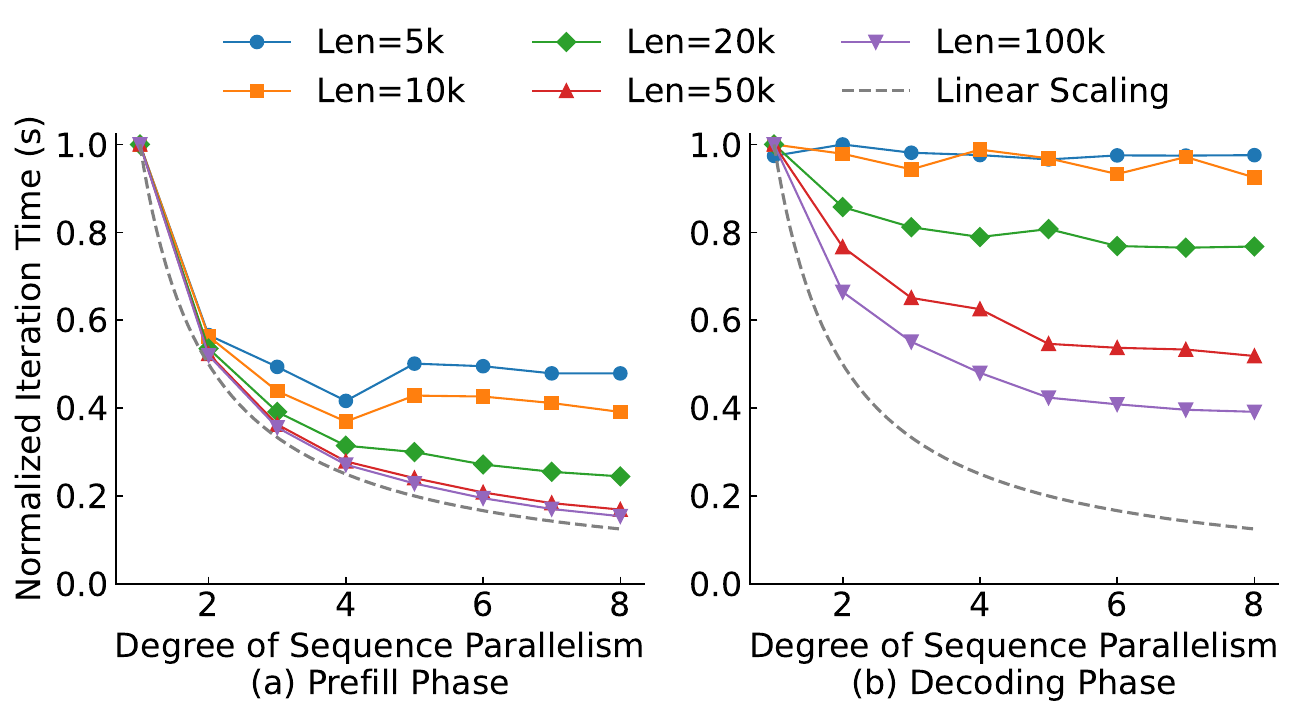}
    \caption{Effectiveness of GPU pooling.}
    \label{fig:background:scalability}
\end{figure}

\parabf{Opportunities.}
Recent advancements in communication hardware and software aspects
have enabled the development of declarative cache designs.
\begin{itemize}[leftmargin=*]
    \item \textbf{Hardware.} A significant opportunity arises from the increasingly powerful interconnections and their inefficient usage.
    The scale-out links are narrowing the gap with PCIe~\cite{HostCongestionControl,a100}, and the scale-up links are even surpassing PCIe~\cite{nvl72,shou2025infinitehbdbuildingdatacenterscalehighbandwidth}, while GPUDirect RDMA is also maturing for low-latency communication.
    However, the bandwidth is often underutilized. For instance, in the attention module, inference traffic between instances is typically unidirectional, flowing from a prefill to a decode instance, while much of the interconnection is idle.
    It leaves a large optimization space to improve performance by unleashing the full potential of modern networking.
    \item \textbf{Software.} A further opportunity exists in optimizing the communication patterns of prefix cache management. Figure~\ref{fig:background:scalability} shows an example. Rather than adhering to transmitting only prefix cache, by combining communication with Sequence Parallelism (SP) in the prefill phase or transmitting typically smaller query tensors in the decoding phase, the performance gains from pooling GPU resources across can often outweigh the communication overhead, because they enable less communication volume and overlap communication better with computation.
\end{itemize}

%% file: sections/overview.tex
\section{System Architecture}
\label{sec:overview}

\subsection{Overview}
\begin{figure}[t]
    \includegraphics[width=\linewidth]{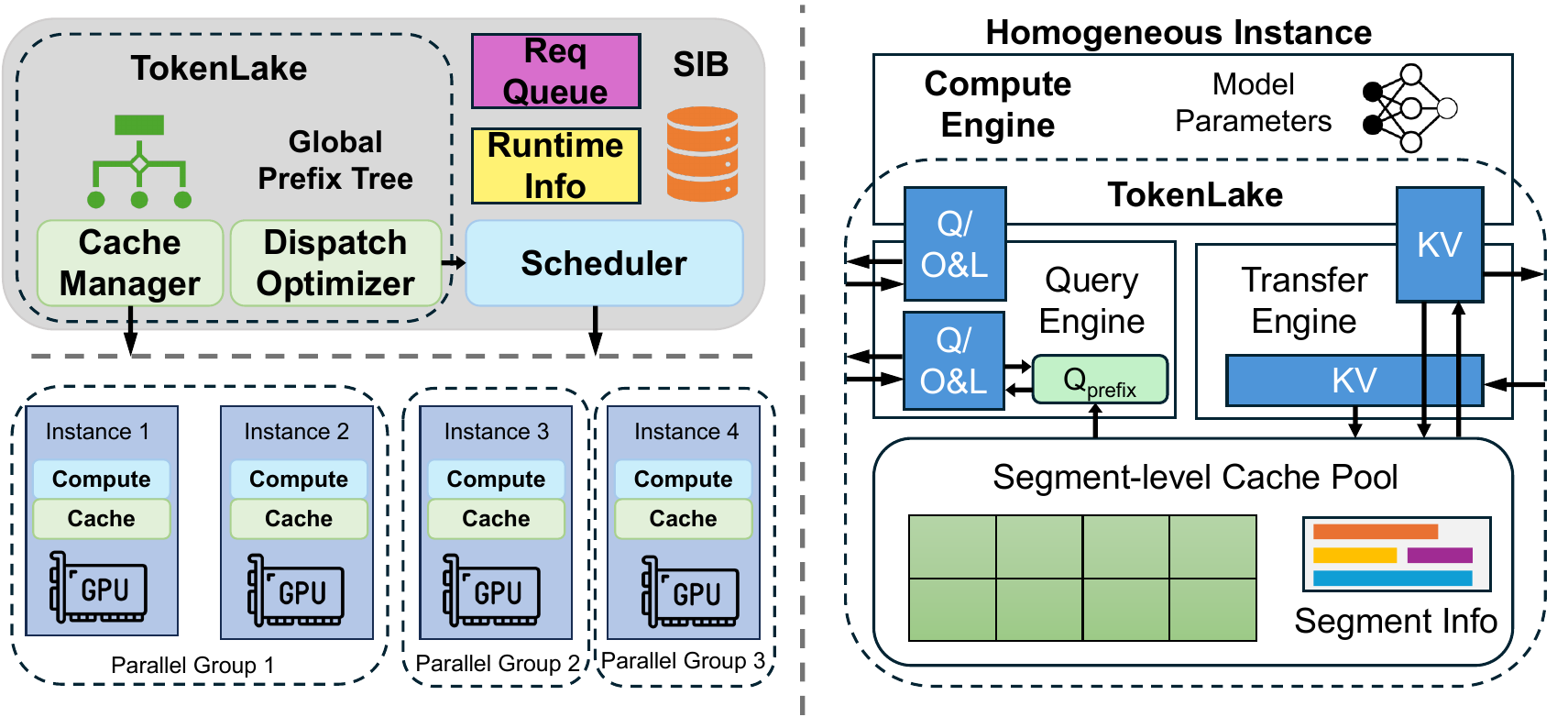}
    \caption{System overview.}
    \label{fig:overview:overview}
\end{figure}

To address these problems, as shown in Figure~\ref{fig:overview:overview}, we design and build a unified segment-level prefix cache pool, \sysname, across instances. \sysname first introduces a declarative prefix cache interface that abstracts cache-aware operations from request scheduling and exposes it to \sysname for better cache management (\S\ref{sec:overview:interface}). Under the hood, prefix caches are split into segments to reduce fragmentation. \sysname uses a fully peer-to-peer asynchronous architecture (\S\ref{sec:overview:p2p}) to enable low-latency asynchronous communication of cache segments and query tensors (\S\ref{sec:overview:ll}). With the global view, \sysname's cache manager uses a heavy-hitter-aware segment-level load balancing algorithm to achieve better load balance and conduct aggressive deduplication (\S\ref{sec:load_balance}) of existing prefix caches. For query tensors and newly generated prefix caches, \sysname's dispatch optimizer minimizes their communication volume with a bipartite matching-based dispatching algorithm (\S\ref{sec:placement}). Finally, stateless elastic scheduling can be performed by the scheduler with minimal consideration of the underlying prefix cache (\S\ref{sec:scheduling}).

\subsection{Interfaces}
\label{sec:overview:interface}

\begin{figure}[t]
    \centering
    \includegraphics[width=\columnwidth]{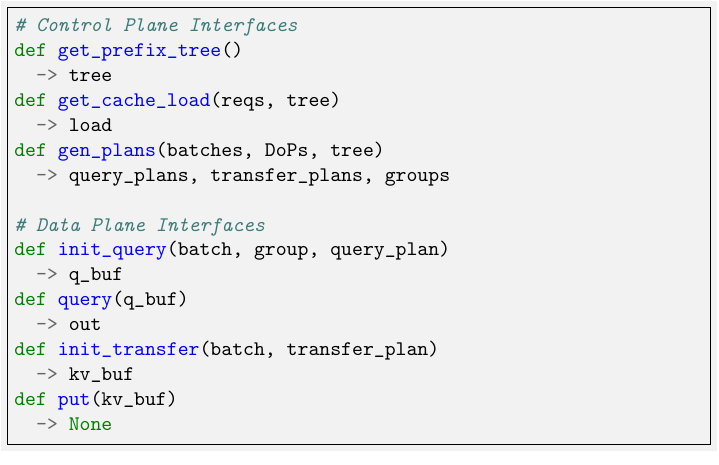}
    \caption{Interfaces of \sysname}
    \label{fig:overview:interface}
\end{figure}

Figure~\ref{fig:overview:interface} shows the declarative interfaces of \sysname. At the control plane, \texttt{get\_prefix\_tree} provides the global prefix tree to the scheduler. The Scheduler can leverage it to get caching information, e.g., hit rate of each request, slots utilization, distribution of cache segments, etc, similar to a local prefix tree in prior works~\cite{sglang}. \texttt{get\_cache\_load} provides the GPU load of \sysname given a set of requests \texttt{reqs} with specified input lengths. The scheduler can use this information to avoid interference with \sysname (\S\ref{sec:scheduling}).
In each iteration, after the scheduler decides batching \texttt{batches} and their respective degree of parallelism \texttt{DoPs}, \texttt{gen\_plans} generates control flow \texttt{query\_plans} and \texttt{transfer\_plans} for the data plane and allocates \texttt{DoP} instances \texttt{group} to each batch for elastic parallel computing.

At the data plane, the compute engine in each instance can initialize the query engine and transfer engine with \texttt{init\_query} and \texttt{init\_transfer} by respective control flows to get the shared buffers \texttt{q\_buf} and \texttt{kv\_buf} in GPU memory. During inference, the compute engine prepares the query tensor in \texttt{q\_buf} and invokes \texttt{query} to get the attention output on the global prefix cache pool. Similarly, after the compute engine prepares the layer-wise newly generated KV cache in \texttt{kv\_buf}, it can invoke \texttt{put} to store them in \sysname asynchronously. Due to the declarative nature of these interfaces, the compute engine does not need to know how to store and query the prefix cache. \sysname automatically optimizes the prefix cache pool and query plans to achieve high performance. At the same time, the degree of parallelism, batching of requests, and splitting input sequences are still fully controlled by the scheduler, enabling elastic scheduling.

\subsection{Fully peer-to-peer asynchronous architecture}
\label{sec:overview:p2p}

To support fine-grained co-optimization of query tensors, prefix cache segments, and prefix attention, \sysname employs a fully peer-to-peer (P2P) asynchronous architecture. As shown in Figure~\ref{fig:overview:overview}, the two main components of \sysname: the query engine and the transfer engine, run as independent processes, colocated on the GPU with the compute engine. Powered by NVIDIA Multi-Process Service (MPS)~\cite{mps}, three engines can run asynchronously and concurrently. They can even handle different tasks for different batches concurrently for better performance.

Query and transfer engines themselves also support multitasking. Query engines are responsible for generating partial outputs and normalizers of the query on local prefix segments. They use multiple CUDA streams to concurrently send query tensors to other query engines and handle query tensors from other query engines concurrently. Similarly, transfer engines can concurrently transfer KV cache to or from different instances. This design maximizes the utilization of interconnect bandwidth.

\subsection{Low-latency asynchronous communication}
\label{sec:overview:ll}

Powered by the fully P2P asynchronous architecture, it is possible to achieve low-latency asynchronous communication for both query tensors and prefix caches.

First, \sysname enables zero-copy data transfer between engines. \texttt{q\_buf} and \texttt{kv\_buf} are shared buffers allocated with CUDA IPC~\cite{cuda}. As long as the compute engine generates tensors directly in these buffers, no extra GPU memory copy is needed between engines. Furthermore, these buffers are also registered with \texttt{ncclCommRegister} as network buffers, minimizing memory copies during network transfers.

\begin{figure}[t]
    \includegraphics[width=\linewidth]{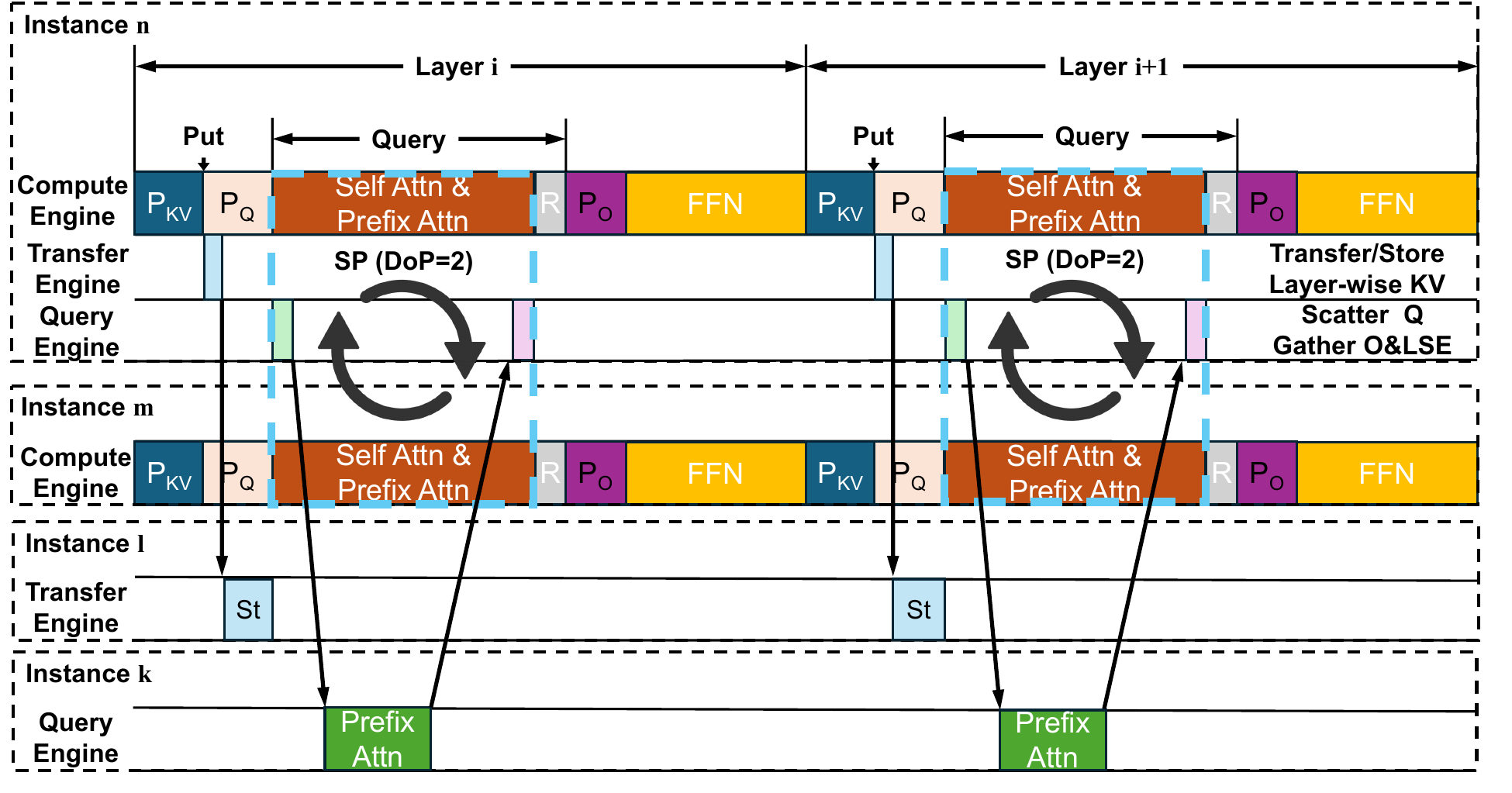}
    \caption{Interactions of instance $n$ with remote engines.}
    \label{fig:design:timeline}
\end{figure}

Second, communication can be overlapped with computation. As shown in Figure~\ref{fig:design:timeline}, layer-wise KV cache transfer ($Put$) can be overlapped with the subsequent computation of the entire layer after generation ($P_{kv}$). Similarly, scattering query tensors and gathering query results, whose communication volume is linear to the sequence length, can be overlapped with the self-attention (\textit{Self Attn}), which is quadratic to the sequence length.

Third, the communication interference is minimized. Different engines are triggered at different times. The transfer engine is triggered after $P_{kv}$, while the query engine is triggered after $P_Q$. Furthermore, if the scheduler decides to use sequence parallelism (SP) to perform self-attention across instances, $Query$ will use the pass-Q variant of SP~\cite{yang2024context} that swaps query tensors within the parallel group in the ring style and removes redundant transmission of query tensors for attending to the prefix cache in the SP group. Therefore, the destined instances of the query engine are different from the instances involving SP, reducing network contention.

Last but not least, the communication volume is small. For example, for a Llama-7B model in its decoding phase, the KV cache transfer at each layer is only 16 KB per request. The query tensor communication volume is also small, which is approximately equal to $2dlN_p$, where $d$ is hidden size, $l$ is the sequence length, and $N_p$ is the number of instances hosting relevant segments. Compared to the communication in large-scale Expert Parallelism (EP)~\cite{zhao2025insights} that possibly occurs in $FFN$, whose volume is equal to $2dlN_e$, where $N_e$ is the number of experts, the query engine's communication volumes are often comparable (\S\ref{sec:load_balance}), representing a widely accepted trade-off in practical deployments. The analysis also aligns with the experimental results in Figure~\ref{fig:background:scalability}.

%% file: sections/design.tex
\section{Segment-based Load Balancing}
\label{sec:load_balance}

Because the declarative interface exposes query tensors and prefix attention to \sysname, prefix cache segments are essentially independent of each other. Query tensors can attend to them independently, generate partial output tensors, and normalizers. The final output can be generated by combining them together~\cite{flashinfer}. Therefore, prefix cache segments can be split and distributed arbitrarily across instances in \sysname. However, even with this flexibility, it is still challenging to efficiently utilize all GPU memories and memory bandwidth as a cache pool.

\subsection{Strawman solution: Strict locality-based policy}
A straightforward solution is to maintain locality whenever possible, which we term a strict locality-based policy. It utilizes other instances only when the local cache capacity is exhausted. However, this strict locality-based policy suffers from two major drawbacks: severe load imbalance and frequent data migration.

\begin{figure}[t]
    \includegraphics[width=\linewidth]{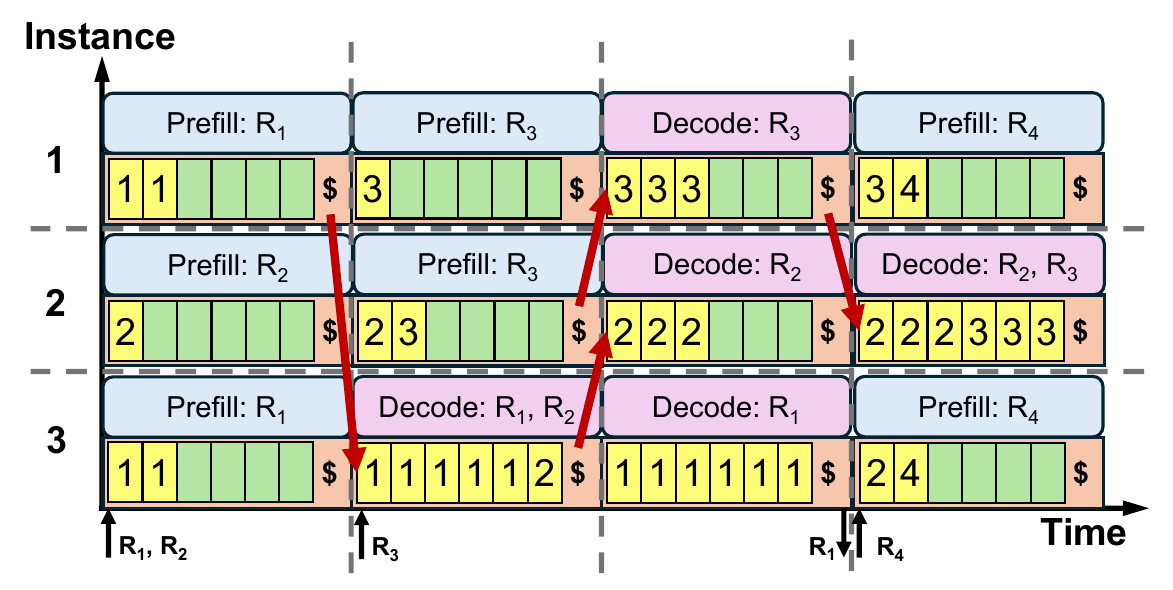}
    \caption{Limitations of strict locality-based policy.}
    \label{fig:design:locality_based_policy}
\end{figure}

First, it ignores memory bandwidth utilization across instances, leading to load imbalance. As shown in Figure~\ref{fig:design:locality_based_policy}, it makes prefix cache segments concentrate on a few instances (e.g., instance 3 in iterations 2 and 3, and instance 2 in iteration 4) with high memory bandwidth pressure, while other instances are under-utilized. This disparity becomes more pronounced as the variation of sequence lengths grows and shared prefix lengths become longer.

Second, the policy reacts poorly to the dynamic workload. Because sequence lengths and phases of requests change over time, the set of instances occupied by a request also needs to change accordingly to maximize overall performance~\cite{wu2024loongserve,patel2024splitwise}. To enforce strict locality, \sysname triggers frequent unnecessary migration to maintain locality. As shown in Figure~\ref{fig:design:locality_based_policy}, to leave computation resources for new requests ($R1$ and $R2$ in iteration 1) or to balance the computation ($R1$, $R2$, and $R3$ in iteration 2), segments need to be migrated frequently to keep strict locality, leading to high communication overhead.

\subsection{Segment size tradeoff: locality vs. load balance}
In contrast, our key insight is that although locality is important to reduce communication volume, it is not strictly necessary for achieving high performance. As long as the segment size is larger than a certain threshold, the distributed query could achieve better performance than the local query. This is because the memory bandwidth and compute of multiple GPUs can be aggregated to process the prefix attention, and the communication overhead can be amortized over the entire segment and overlapped by local self-attention computation. Furthermore, it also avoids load imbalance and frequent data migration caused by strict locality.

Consider a prefix cache of sequence length $S$ partitioned across $N$ instances. In each decoding iteration, a single query tensor $q$ must attend to the entire prefix cache. The cache on each instance represents a data segment of size $C$, where $N=S/C$. The goal is that when the segment resides on a remote instance, the local computation latency saved is greater than or equal to the communication overhead incurred.

The saved computation latency to process a segment of size $C$ is bound by either compute or memory bandwidth according to the Roofline model. Let $F$ be the floating-point operations per second (FLOPS) of a single GPU, $B_{mem}$ be the memory bandwidth, and $d$ be the hidden dimension. The time to process a segment of size $C$ is:
$
T_{comp}(C) = \max(\frac{4Cd}{F}, \frac{4Cd}{B_{mem}}) = k_{comp}C
$,
where $k_{comp} = \max(\frac{4d}{F}, \frac{4d}{B_{mem}})$ is a constant for a given model and hardware.

The communication time consists of two phases, including sending query tensors and receiving partial output tensors and normalizers. In the whole process, the total communication latency is:
$T_{comm}(C) = 2\alpha_{net} + \frac{4d}{B_{net}}$,
where $\alpha_{net}$ is the network latency, and $B_{net}$ is the network bandwidth.

If a segment size can make computation latency saved greater than or equal to the communication latency, then we can derive:
$C \geq \frac{2\alpha_{net}}{k_{comp}} + \frac{4d}{k_{comp}B_{net}}$.
So if we set the segment size $C$ to be larger than this threshold, it can harvest memory bandwidth and compute on remote instances without incurring significant communication overhead.
When we instantiate this threshold using parameters from our experimental setup (\S~\ref{sec:evaluation}), $C$ is approximately equal to 568 tokens, enabling fine-grained segment-level cache pooling possible. Furthermore, even if the communication overhead can not be completely hidden, it is still negligible compared to the overall computation time. For example, based on the above analysis, when $C$ is 568 tokens, the communication overhead of a short request in the decoding phase with 1000 prefix tokens is only 4.66 us per layer, a negligible time in practice.

\subsection{Heavy-hitter-aware segment-level load balancing}
After choosing a segment size larger than the threshold, the next challenge is how to distribute them to achieve a balanced load across instances. Highly skewed and dynamic access patterns make it challenging. Because a segment is accessed only when the prefix of this segment is accessed, the access pattern is inherently skewed, concentrating on segments closer to the root of the prefix tree. Furthermore, because the workload is dynamic, the access pattern changes over time. The rapid processing speed exacerbates this issue.

To address this problem, we propose a heavy-hitter-aware load-balancing algorithm that handles heavy hitters and normal segments differently. Heavy hitter is defined as frequently accessed segments. For normal segments, \sysname employs a hash-based distribution strategy to spread them uniformly across all instances. A hash function is used to map each segment to an instance. Given that normal segments are numerous, this method typically achieves effective load balancing without migrations and aggressive deduplication.

As for heavy hitters, \sysname employs a selective replication strategy to maintain load balance. If an instance is overloaded,
it replicates heavy hitter segments on that instance to the least-loaded instances without its copy. Because heavy hitters reside closer to the prefix tree's root, \sysname can find them by using a breadth-first search (BFS) without traversing the entire prefix tree.
To make the replication efficiently, our approach is informed by a well-established result from early front-end caching systems~\cite{fan2011small}: \textit{if a system can effectively absorb the traffic from the top $O(N \log N)$ heavy hitters, the load across $N$ storage servers is guaranteed to be balanced with high probability, regardless of the query distribution or the total number of objects}. Therefore, the number of heavy hitters is set to $O(N \log N)$ for quick replication.

\subsection{Replicated segment selection and eviction}
For replicated segments, \sysname uses the power-of-two-choices (PoT) policy to select a replica for matched requests. Specifically, it directs the query to the less loaded of the two randomly sampled replicas. It is a variant of the balls-into-bins problem and has been proven that \textit{it can achieve near-perfect load balancing with high probability if traffic to a segment is less than half of the total traffic}~\cite{distcache}, which always holds in \sysname. For eviction, the global LRU policy is used by recording the last access time of replicas.

\begin{figure}[t]
    \includegraphics[width=\linewidth]{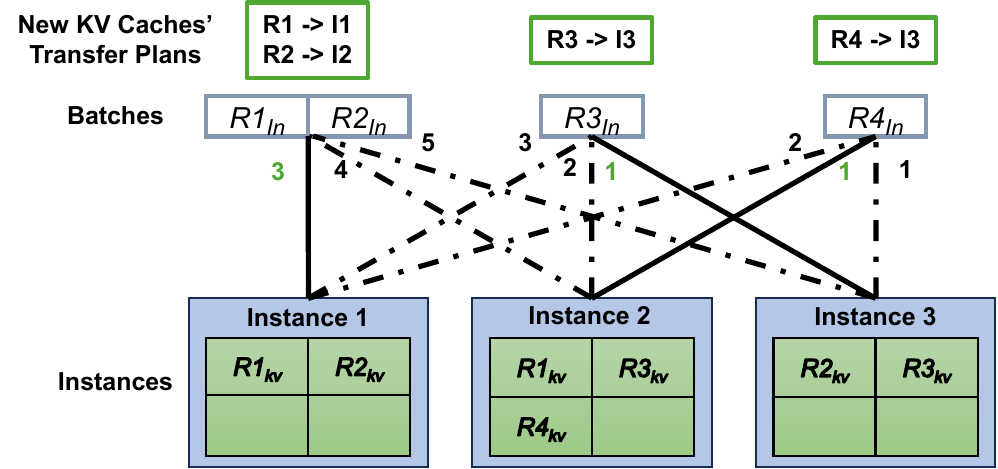}
    \caption{Example of dispatching problem.}
    \label{fig:design:matching}
\end{figure}

\section{Dispatching Optimization}
\label{sec:placement}

\subsection{Problem formulation}
Although the load balancing algorithm can effectively utilize GPU memories and memory bandwidth, the communication volume of \texttt{query} and \texttt{put} operations can still change according to the dispatching of requests.
Specifically, for a batch of requests provided by the scheduler, if segments required by the batch are predominantly located on an instance, or if the newly generated KV caches are primarily intended to be stored on an instance, \sysname can dispatch the batch to that instance to reduce communication volume.

Figure~\ref{fig:design:matching} illustrates an example where the scheduler has grouped four requests into three batches. The communication volume generated by each batch depends on the instance it is placed on. For example, consider the batch containing request $R3$. If it is executed on Instance 3, it only needs to send one query to Instance 2 for attending related segments. Then the communication volume of this dispatching is $1 \times 4d$. If, however, it is executed on Instance 1, a query tensor and the newly generated KV cache need to be sent to Instance 2 for \texttt{query} and Instance 3 for \texttt{put}, leading to $2 \times 4d$ communication volume. Therefore, for efficient memory pooling, the dispatching of batches should be optimized to minimize the overall communication volume.

\subsection{Bipartite matching-based dispatching algorithm}
Regarding this problem, our key insight is that the dispatching can be decoupled from the computation scheduling. Because load has been actively balanced across instances (\S\ref{sec:load_balance}), where to place the computation has no significant impact on the computation. 
Therefore, dispatching can be optimized as an independent problem and be transparent to the scheduler. The scheduler only needs to invoke \texttt{gen\_plans} with batching specifications and their respective DoP. \sysname can optimize dispatching plans transparently. 

For this problem, our key observation is that it can be converted into a bipartite matching problem to efficiently find the solution with the minimum communication volume.
Specifically, we construct a bipartite graph $G=(U,V,E)$, where $u_i \in U$ represents a batch with its transfer plan, and $V$ is the set of instances. Let $Q(u_i)$ be the set of instances caching segments related to $u_i$, $P(u_i)$ be the set of instances where new KV cache of $u_i$ will be cached and $N_p(u_i, v_k)$ is the number of new KV cache of $u_i$ sent to instance $v_k$, according to the transfer plans. For a batch with DoP greater than one, the corresponding node $u_i$ needs to be decomposed into $DoP$ sub-batch nodes $u_{i1}' \cdots u_{id}'$, and the above functions are then defined for each new sub-batch node according to the actual communication of that sub-batch. Then the weight of an edge from $u_i \in U$ to $v_j \in V$ is equal to: 
\[
e(u_i, v_j) = -\left(\sum_{v_k \neq v_j, v_k \in Q(u_i)} 4d + \sum_{v_k \neq v_j, v_k \in P(u_i)} 4d * N_p(u_i, v_k)\right)
\]

With this construction, we convert the problem of minimizing communication volume to finding a perfect matching in $G$ with minimal edge weights. It can be solved by using the Hungarian algorithm~\cite{Hungarian}. The time complexity of this algorithm is $O(n^3)$, where $n$ is the number of instances. Therefore, \sysname can efficiently find the dispatching plan with the minimal communication volume by invoking this algorithm at the beginning of each iteration.

\section{Stateless Elastic Scheduling}
\label{sec:scheduling}

\subsection{Cache load isolation}
With \sysname, the scheduler can still serve for different objectives and combine various scheduling strategies, such as PD disaggregation~\cite{patel2024splitwise}, chunked prefill~\cite{holmes2024deepspeed}, and elastic sequence parallelism~\cite{wu2024loongserve}, as usual, and does not need to manage prefix cache. The only difference is that \sysname inevitably occupies some GPU resources for prefix cache management. But as long as the scheduler is aware of this load, it can adjust its scheduling accordingly to avoid interference. For example, it can reduce the batch size, split long sequences into chunks, or use sequence parallelism to reduce the load on each instance. Because the scheduling becomes stateless, these adjustments can be made efficiently.

In the end, \sysname provides a \texttt{get\_cache\_load} API to expose its own resource consumption. Specifically, \sysname calculates the percentage of GPU resources consumed by \sysname for requests (\texttt{reqs}) in the worst case. First, it estimates the minimum execution time of \texttt{reqs}, $T_r$, under an ideal scenario where all GPUs are fully utilized. This estimation is based on pre-profiled results and follows the quadratic curve fitting method as prior work~\cite{wu2024loongserve}. Next, the memory bandwidth and computation consumption $M_r$ and $F_r$ of \sysname are respectively calculated as follows:
\[
M_r = \sum_{r \in reqs} 4 \times d \times (r.prefix\_len + r.input\_len)
\]
\[
F_r = \sum_{r \in reqs} 2 \times d \times r.prefix\_len \times r.input\_len
\]
Based on the roofline model, the percentage of GPU resources consumed by \sysname is then calculated as:
\[
L = \max(\frac{M_r}{M_r + N \times B_{mem} \times T_r}, \frac{F_r}{F_r + N \times F \times T_r})
\]
Because the load of \sysname is balanced (\S\ref{sec:load_balance}), $L$ is independent of instances.
With this information, the scheduler can factor the load of \sysname into its scheduling decisions.

\subsection{Goodput-optimized stateless elastic scheduling}
To showcase the flexibility and potential of elastic scheduling on \sysname, we designed a goodput-optimized stateless elastic scheduling algorithm. To optimize the overall goodput for requests with varying lengths and in different phases, our optimization objective is to minimize the overall latency, subject to the constraint that the output token latency, a.k.a Time Between Tokens (TBT), meets a Service Level Objective (SLO).
This objective first minimizes the overall processing time. Second, it allows more requests in the prefill phase to enter the decoding phase quickly. This, in turn, increases the batch size during the decoding phase, thereby improving overall GPU efficiency.

To tackle this optimization problem, in each iteration, following existing systems~\cite{sglang,vllm,holmes2024deepspeed}, we first use chunked prefill to split the long context to avoid overloading the whole cluster. Then it can be formulated as a dynamic programming (DP) problem to decide batching and DoP of each batch. Since requests with similar context lengths exhibit similar characteristics, we first sort all requests by their context length. Let $f(i,k)$ represent the minimum overall latency for the first $i$ requests given $k$ instances. The DP equation can be formulated as follows:
\[
f(i,k) = \min_{0 \leq j < i, 0 \leq l < k, \atop SLO_{j+1 ... i} \leq T(j+1, i, k-l, L)} f(j,l) + (i-j) \times T(j+1, i, k-l, L)
\]
where $T(j+1, i, k-l)$ is the latency to batch requests from $j+1$ to $i$ using $k-l$ instances when cache load is $L$. The time is also estimated by the quadratic curve fitting method~\cite{wu2024loongserve}. Then the scheduling plan is constructed by backtracking from $\min_{k \leq N} f(M, k)$, where $M$ is the total number of requests. To ensure PD disaggregation, we constrain requests in a batch to be in the same phase. If the SLO actually cannot be met, the algorithm will fall back to throughput-oriented scheduling by ignoring the SLO constraint to mitigate head-of-line blocking. Because DoP of request in the decoding phase must be one, the algorithm complexity is $O(M_p^2 \times N^2)$, where $M_p$ is the number of requests in the prefill phase.

%% file: sections/evaluation.tex
\begin{figure*}[t]
    \includegraphics[width=\linewidth]{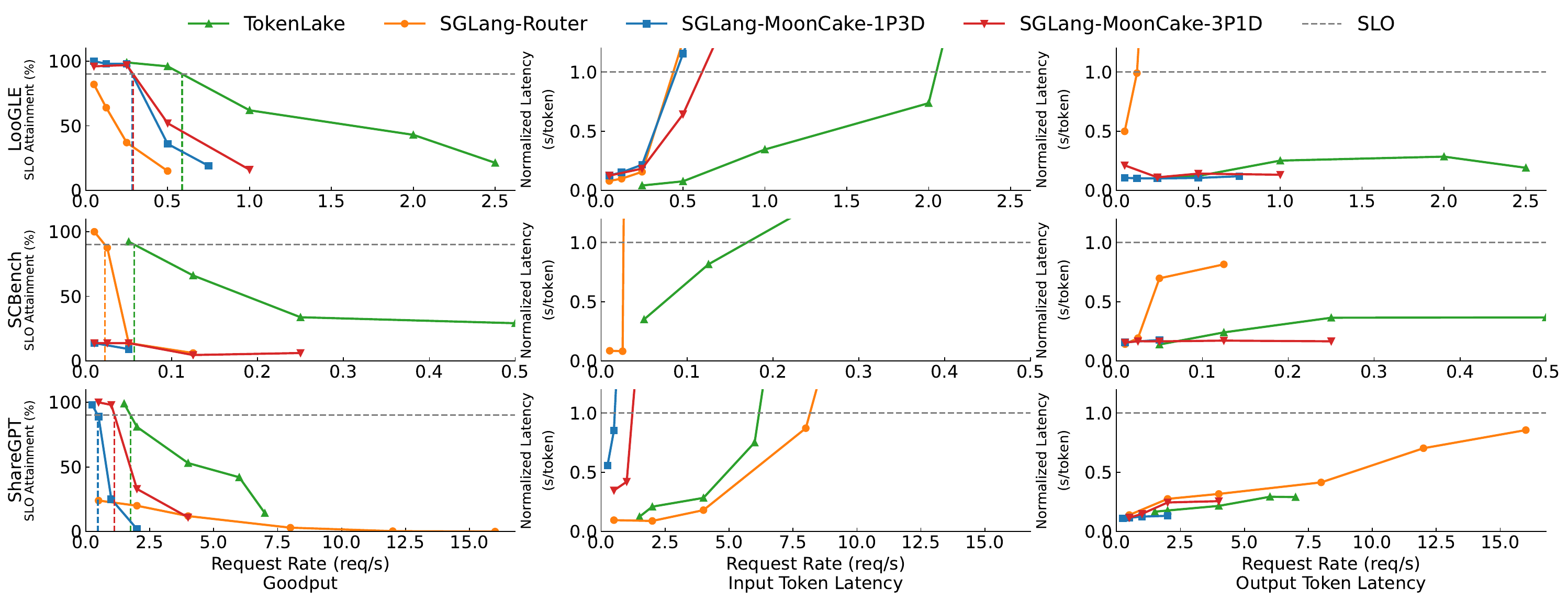}
    \caption{End-to-end performance.}
    \label{fig:evaluation:e2e}
\end{figure*}

\section{Evaluation}
\label{sec:evaluation}

We implement \sysname as a prototype based on FlashInfer~\cite{flashinfer}, SGLang~\cite{sglang}, ZMQ~\cite{zmq}, Ray~\cite{ray}, and LoongServe~\cite{wu2024loongserve} in 16k LoCs. \sysname uses CUDA IPC to share GPU memory among engines and uses inter-process CUDA events and Python semaphores for interactions between the engines. In this section, we evaluate it with state-of-the-art systems on real-world workloads to show its effectiveness.

\textbf{Experimental Setup.} We conduct experiments on servers with 8 NVIDIA A100 80 GB GPUs and 128 CPUs. Most of the experiments are conducted on one server, and we evaluate multi-node performance in Section~\ref{sec:evaluation:multi_node}. The NVLink bandwidth is 400 GB/s. The servers are interconnected via four 200 Gbps InfiniBand NICs. We use CUDA 12.4 and NCCL 2.24.3. Similar to prior works~\cite{wu2024loongserve,liang2025injecting}, we use the LWM-1M-Text, a long-context Llama-2-7B model~\cite{touvron2023llama2openfoundation} for experiments.

\textbf{Workloads}
As prior works~\cite{vllm,zhong2024distserve,wu2024loongserve}, the requests are sampled from the three real-world datasets with the arrival pattern generated by a Poisson process. \textit{LooGLE}~\cite{li-etal-2024-loogle} is a popular dataset of requests with shared prefixes. The average sequence length is 24K. \textit{SCBench}~\cite{li2025scbench} is a dataset of long-context multi-turn sessions. The average sequence length is 227K, and the average number of turns is 5. \textit{ShareGPT}~\cite{sharegpt} is a conversation dataset with short requests. The sequence length is less than 2.4K.
To mimic the real-world scenario, we add a system prompt of OpenAI o3~\cite{o3systemprompt} to each request.

\textbf{Baselines.} vLLM~\cite{vllm} and SGLang~\cite{sglang} are two state-of-the-art open-sourced LLM serving ecosystems without a significant difference. Because we reuse some components from SGLang, we choose solutions on top of it to ensure a fair comparison. Specifically, \textit{SGLang-MoonCake-xPyD} is a cache-centric PD disaggregation solution based on MoonCake~\cite{mooncake} and SGLang. Because the P:D ratio can affect performance, we evaluate two configurations, P:D = 1:3 and P:D = 3:1. \textit{SGLang-Router} is a cache-aware routing solution by combining multiple SGLang instances and an SGL-router. It considers both cache hit rate and load balancing to route requests. Similar to prior works~\cite{wu2024loongserve,cui2025optimizing}, tensor parallelism (TP) is set to 2, and chunked prefill with the same chunked prefill size is enabled.

\textbf{Metrics.} For prefill performance, we measure the input token latency~\cite{lin2025bullet,wu2024loongserve}, i.e., time to first token (TTFT) normalized by the input length. For decoding performance, we measure the output token latency, i.e. the time between tokens (TBT). For the SLO attainment, we set the SLO to be that all input and output token latency is less than 10$\times$ the processing time when the batch size is 1. The P90 goodput is defined as the maximum throughput of the system when 90\% of requests meet the SLO.

\subsection{End-to-end performance}
\label{sec:evaluation:e2e}
We compare \sysname against three distinct baselines across three different datasets on three key metrics. As shown in Figure~\ref{fig:evaluation:e2e}, \sysname outperforms the other baselines on nearly all datasets, particularly in SLO attainment.

The first row shows the results on LooGLE.
For SGLang-Router, 
although it attempts to balance computation load, it still tends to direct requests to instances with related prefixes. Given the limited local cache size, this often results in cache thrashing and costly recomputation. Consequently, at the same input token latency, its throughput is 4.64$\times$ lower than that of \sysname. Its output token latency is also high due to the interference from prefill requests.
In contrast, SGLang-MoonCake-1P-3D and SGLang-MoonCake-3P-1D, although protecting the decoding phase, only 25\% and 75\% of prefix cache slots can be used during the prefill phase, respectively, leading to a lower cache hit rate. Furthermore, because only 75\% and 25\% of prefix cache slots can be used during the decoding phase, requests in the prefill phase have to wait for cache slots in decode instances to be released, which also inflates the input token latency.
By efficiently pooling GPU memories, \sysname improves P90 goodput by 2.08$\times$ and 2.04$\times$ over SGLang-MoonCake-1P-3D and SGLang-MoonCake-3P-1D, respectively. At equivalent input token latencies, the throughput improvements are 4.46$\times$ and 3.32$\times$.

\begin{figure*}[t]
    \centering
    \begin{minipage}{0.24\linewidth}
        \centering
        \includegraphics[width=\linewidth]{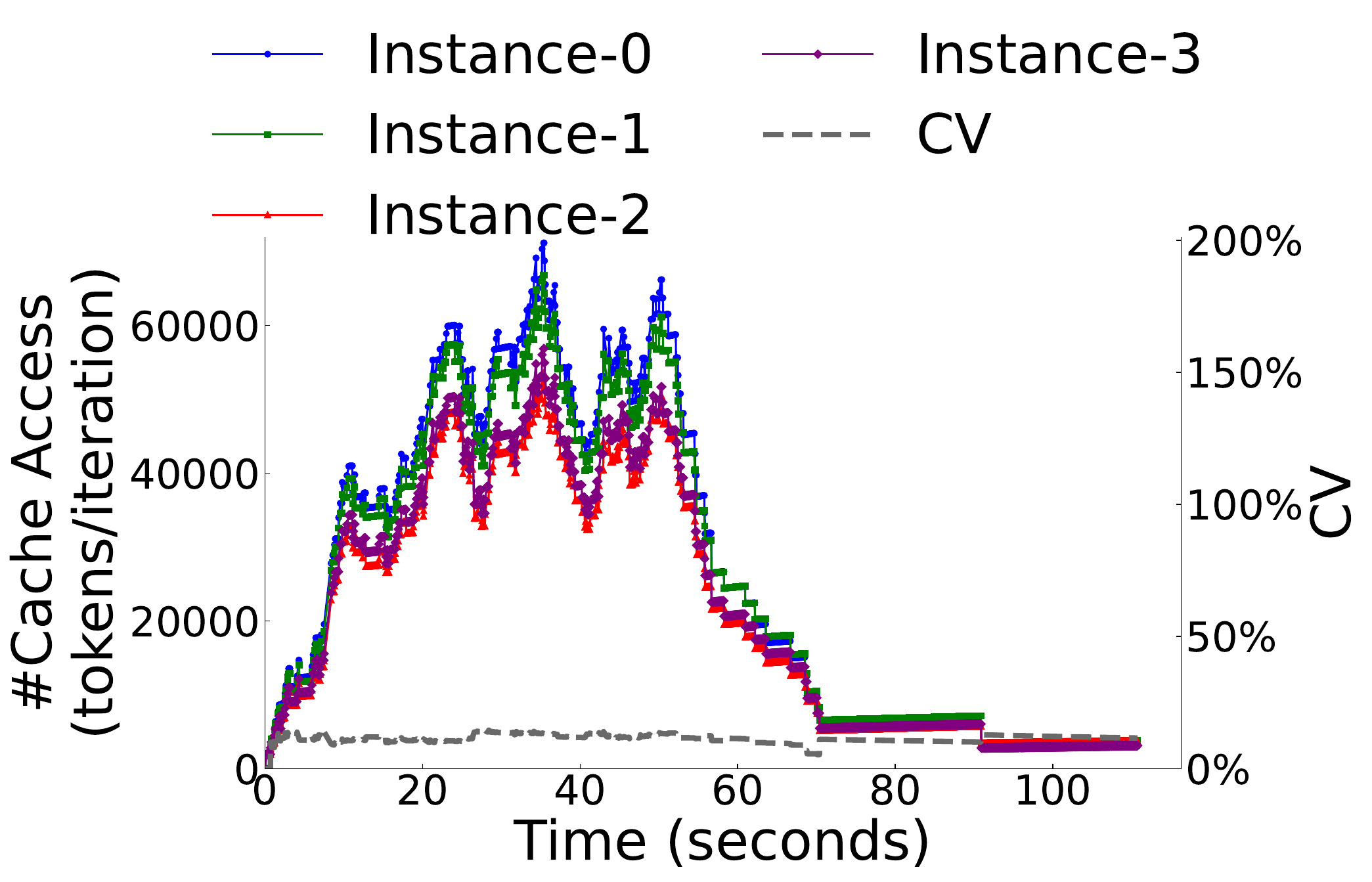}
        \subcaption{\sysname}
        \label{fig:times:tokenlake}
    \end{minipage}
    \hfill
    \begin{minipage}{0.24\linewidth}
        \centering
        \includegraphics[width=\linewidth]{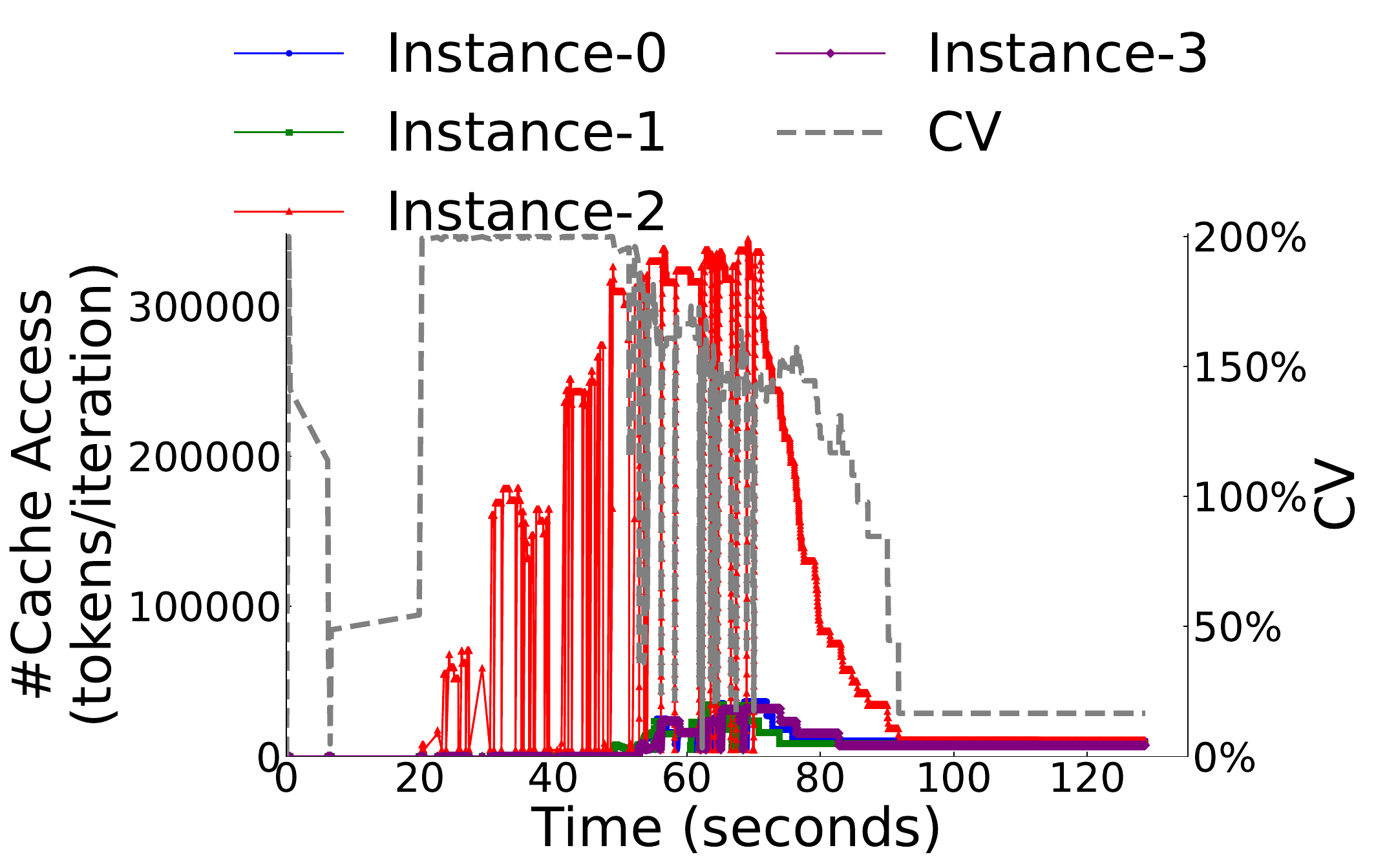}
        \subcaption{SGLang-Router}
        \label{fig:times:sgl_router}
    \end{minipage}
    \hfill
    \begin{minipage}{0.24\linewidth}
        \centering
        \includegraphics[width=\linewidth]{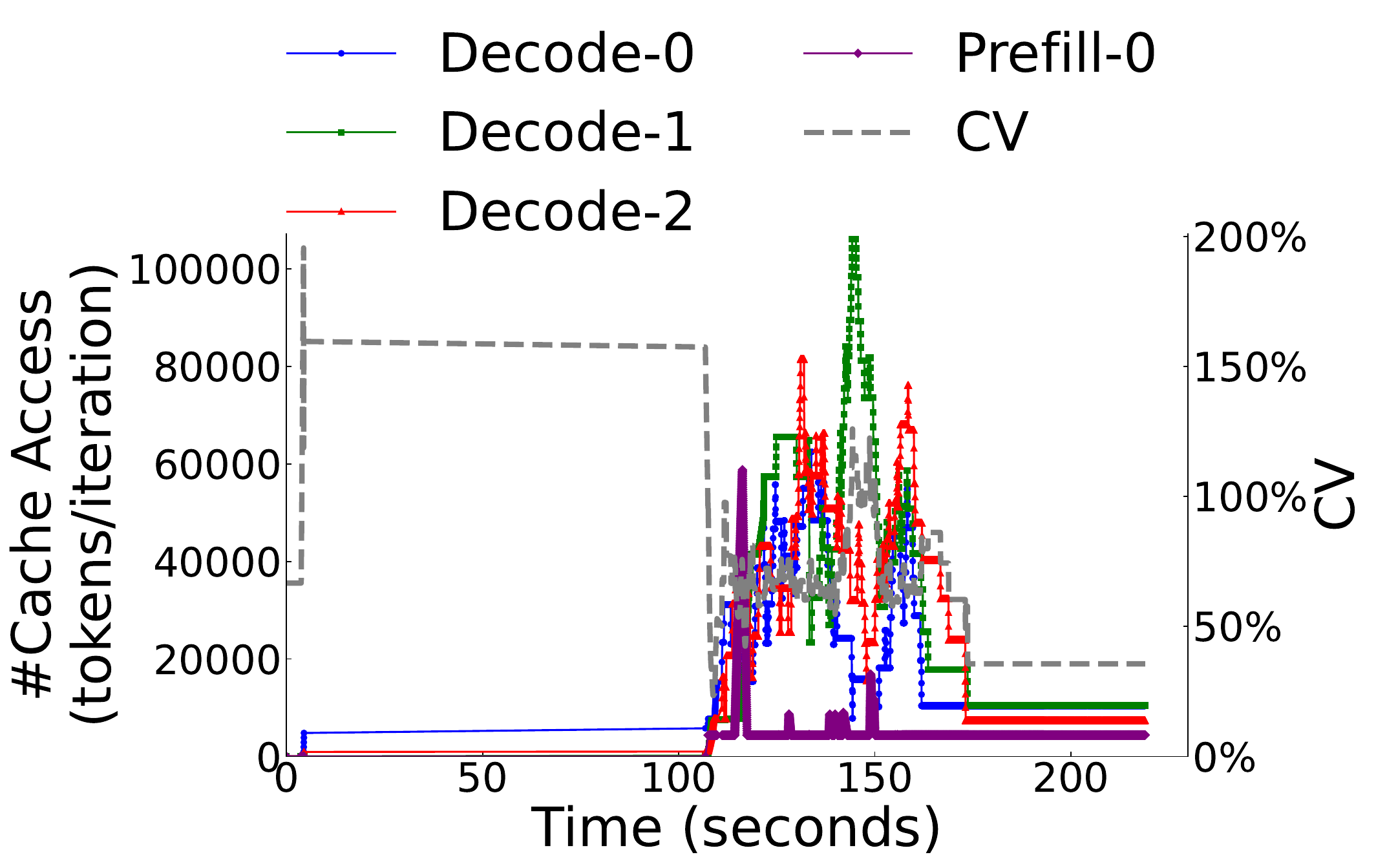}
        \subcaption{SGLang-MoonCake-1P3D}
        \label{fig:times:sgl_mooncake_1p_3d}
    \end{minipage}
    \hfill
    \begin{minipage}{0.24\linewidth}
        \centering
        \includegraphics[width=\linewidth]{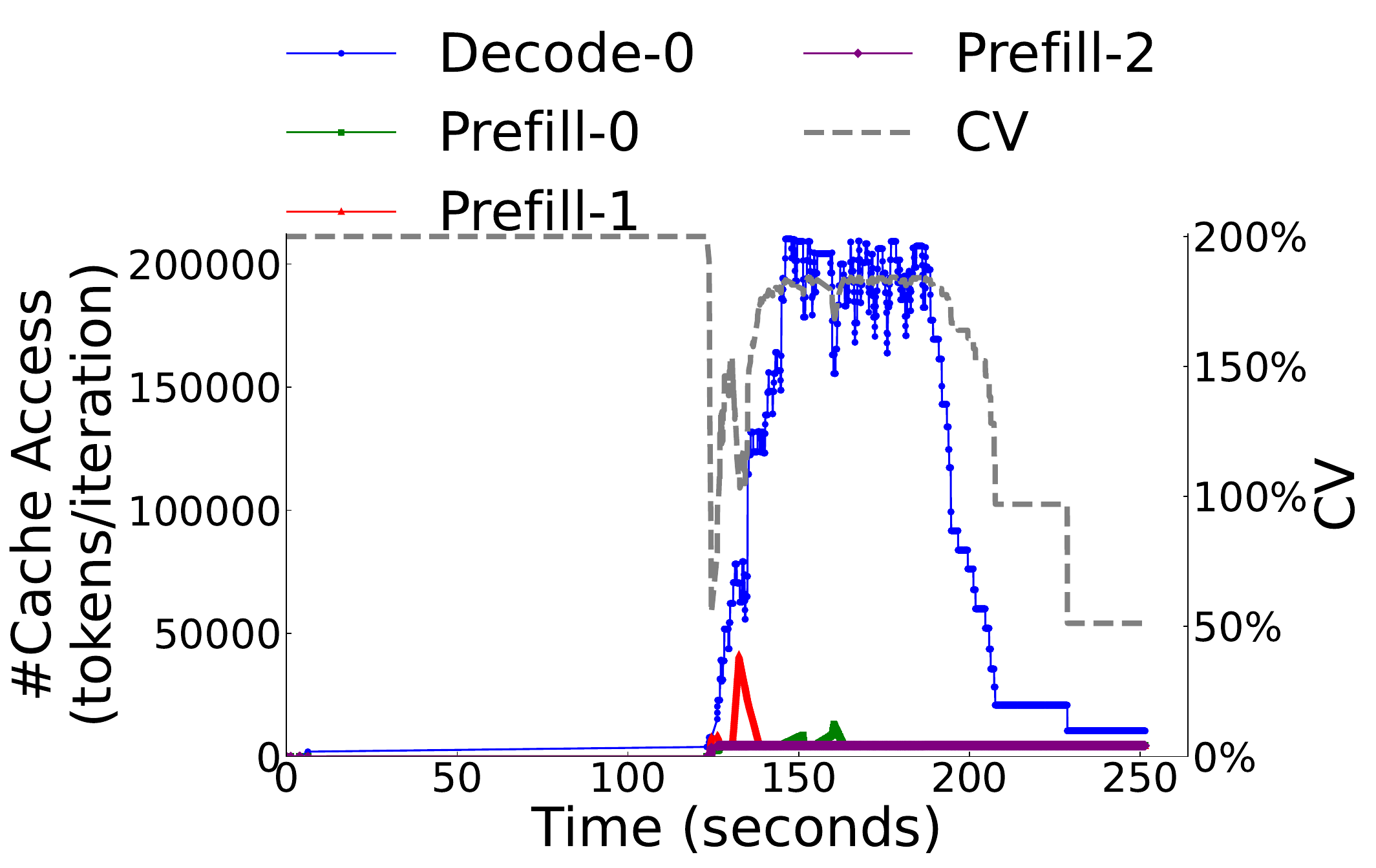}
        \subcaption{SGLang-MoonCake-3P1D}
        \label{fig:times:sgl_mooncake_3p_1d}
    \end{minipage}
    \caption{Load balancing of different systems.}
    \label{fig:times:comparison}
\end{figure*}

\begin{figure}[t]
    \includegraphics[width=\linewidth]{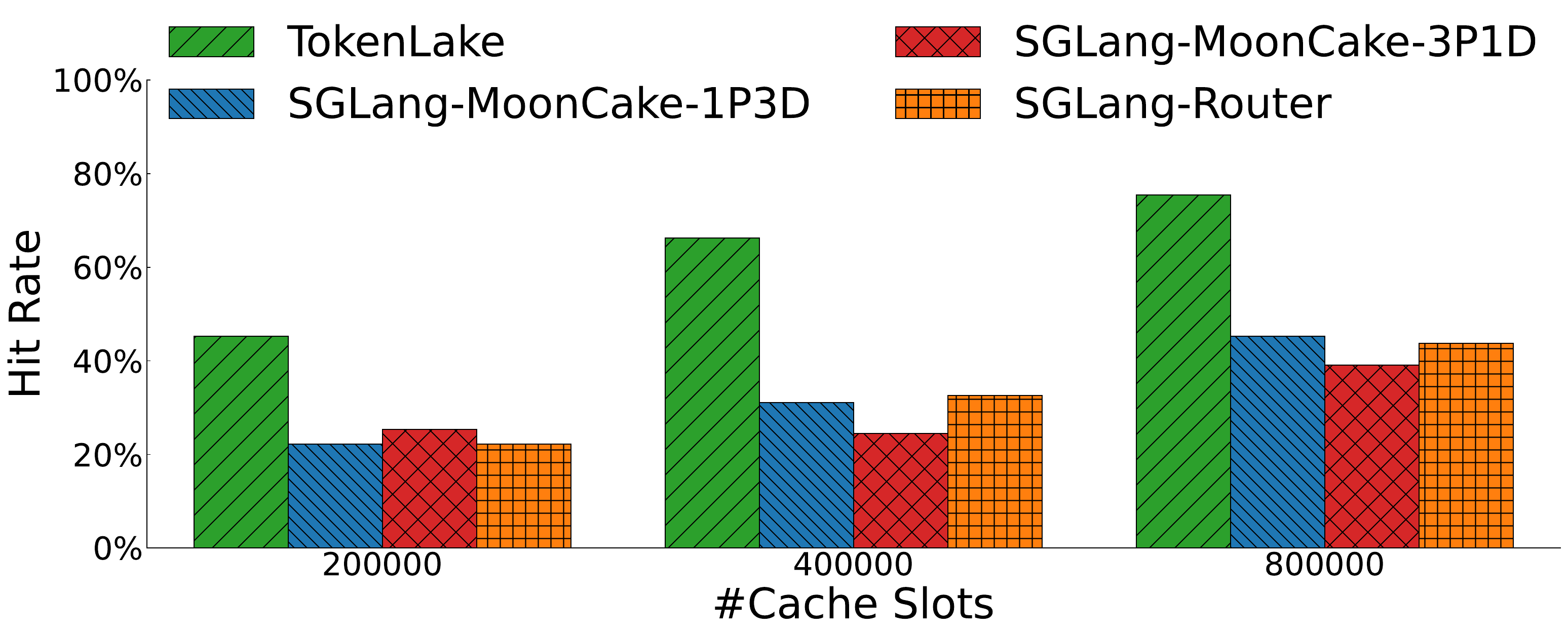}
    \caption{Hit rate under different cache sizes.}
    \label{fig:evaluation:hit_rate}
\end{figure}

On SCBench, because it contains multi-turn conversation, which involves multiple cycles of prefill and decoding phases, cache-centric PD disaggregation has to frequently transfer prefix caches between prefill and decode instances, creating substantial transfer overhead. Fragmentation between two phases also decreases the hit rate. Therefore, SGLang-MoonCake-1P-3D and SGLang-MoonCake-3P-1D can only serve about 20\% of requests, essentially the first turn of a conversation, under the SLO.
Although SGLang-Router does not suffer from this issue, it still causes a low hit rate due to inherent fragmentation and load imbalance across instances.
Consequently, \sysname achieves 2.55$\times$ higher P90 goodput than SGLang-Router. Moreover, because \sysname employs PD disaggregation, its decoding phase latency is also significantly lower than SGLang-Router's.

Finally, on the ShareGPT dataset, \sysname demonstrates the flexibility of its interfaces. Different from SGLang-Router, which sacrifices the SLO attainment, it prevents interference between the two phases as PD disaggregation does, while empowering the dynamic adjustments of the resources allocated to the two phases based on real-time workload. As a result, compared to SGLang-MoonCake-1P-3D and SGLang-MoonCake-3P-1D with the static partitioning, \sysname improves the P90 goodput by 3.71$\times$ and 1.56$\times$ respectively.

\subsection{Load balancing comparison}

To compare the load-balancing capability, we monitored the prefix cache accesses on each instance at a request rate approaching \sysname's P90 SLO threshold. Figure~\ref{fig:times:comparison} shows both the per-instance cache access pattern over time and the corresponding Coefficient of Variation (CV), which serves to quantify the degree of load imbalance across instances.

As shown in Figure~\ref{fig:times:tokenlake}, the CV of \sysname remains steadily below 15\%, with an average value of just 11\%. This result demonstrates the effectiveness of \sysname's load-balancing mechanism over dynamic workloads.

As shown in Figure~\ref{fig:times:sgl_router}, SGLang-Router exhibits significant shortcomings in load balancing. Although SGLang Router takes load conditions into account when routing requests, without memory pooling, it has to trade load balance for a higher hit rate.
Therefore, its average CV reached 99\%, substantially 9$\times$ higher than that of \sysname.

As shown in Figure~\ref{fig:times:sgl_mooncake_1p_3d} and Figure~\ref{fig:times:sgl_mooncake_3p_1d}, due to disaggregation, both SGLang-MoonCake-1P3D and SGLang-MoonCake-3P1D exhibit significant load imbalance. Furthermore, the load imbalance also occurs across decode instances of SGLang-MoonCake-1P3D and across prefill instances of SGLang-MoonCake-3P1D. The figure also reveals that without memory pooling, they serve initial requests at a low hit rate. Ultimately, their average CV reaches 62\% and 122\%, respectively, 5.63$\times$ and 11.09$\times$ higher than that of \sysname.

\begin{figure}[t]
    \includegraphics[width=\linewidth]{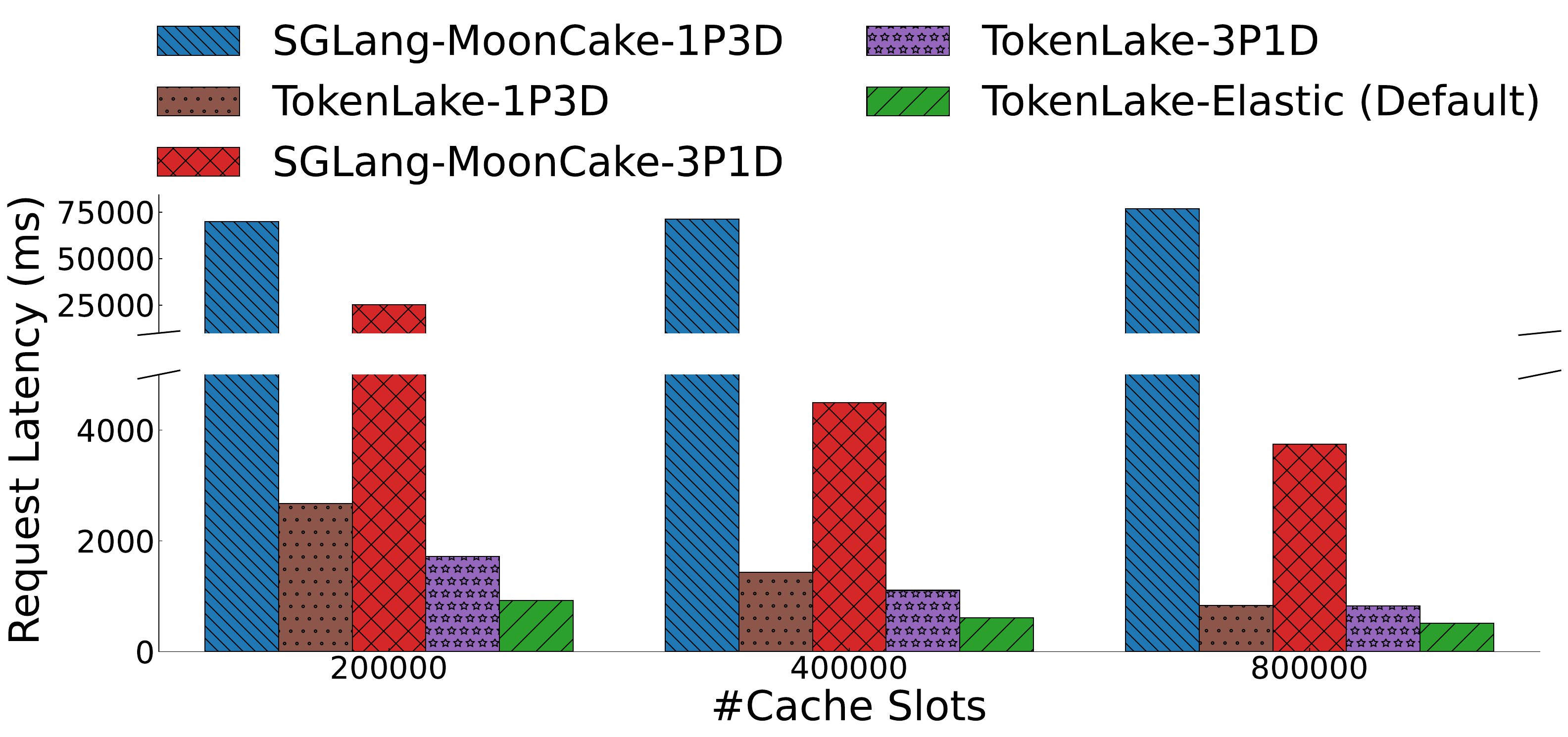}
    \caption{Effectiveness of pooling and elasticity.}
    \label{fig:evaluation:elastic_scheduling}
\end{figure}

\begin{figure*}[t]
    \includegraphics[width=\linewidth]{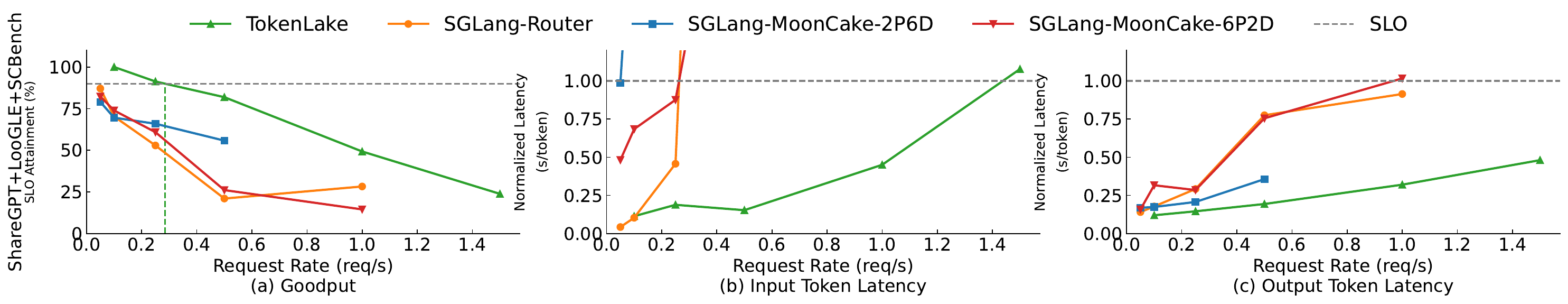}
    \caption{Multi-node performance.}
    \label{fig:evaluation:multi_node}
\end{figure*}

\subsection{Hit rate analysis}
We then analyzed the hit rate of each system under varying cache sizes using the LooGLE dataset. As shown in Figure~\ref{fig:evaluation:hit_rate}, the hit rate of \sysname is the highest across all cache sizes and consistently improves as the capacity increases.
In contrast, the SGLang-MoonCake-1P3D and SGLang-MoonCake-3P1D suffer from their isolated prefix caches for the prefill and decoding phases, so their hit rates are 1.67-2.14$\times$ and 1.78-2.70$\times$ lower than \sysname's, respectively.
SGLang-Router also exhibits a low hit rate due to the limitations of its isolated local caches, which lead to load imbalance, data redundancy, and memory fragmentation. Furthermore, the interference from the requests in the prefill phase slows down its decoding phase performance, which in turn diminishes cache utilization efficiency. Therefore, the hit rate of \sysname is 1.72-2.04$\times$ higher than SGLang-Router's.

\subsection{Effectiveness of pooling and elasticity}
To evaluate the respective effectiveness of the \sysname interface to support elastic scheduling and prefix cache pooling, we implemented PD disaggregation (\textit{\sysname-1P3D} and \textit{\sysname-3P1D}) on top of \sysname to isolate their contributions.
As shown in Figure~\ref{fig:evaluation:elastic_scheduling}, first, the latency of \sysname-1P3D and \sysname-3P1D is 26.62-86.05$\times$ and 2.18-39.48$\times$ lower than that of SGLang-MoonCake-1P3D and SGLang-MoonCake-3P1D, respectively, showing the effectiveness of prefix cache pooling.
Second, our fully-featured system, \sysname-Elastic, consistently outperforms other \sysname variants across all cache sizes. The performance advantage becomes more pronounced as the cache size decreases, reducing latency by up to 5.22$\times$ and 2.82$\times$ compared to \sysname-1P3D and \sysname-3P1D, respectively. This result validates the effectiveness of the \sysname interface to enable stateless elastic scheduling.

\subsection{Multi-node performance}
\label{sec:evaluation:multi_node}

Finally, we also tested the multi-node performance of \sysname on 16 A800 80GB GPUs across two servers. We mixed the three datasets in a 1:1:1 ratio as a more dynamic workload. The PD disaggregation baselines are adjusted to SGLang-MoonCake-6P2D and SGLang-MoonCake-2P6D to maintain the original PD ratios. As shown in Figure~\ref{fig:evaluation:multi_node}, \sysname still outperforms the baselines in multi-node scenarios. Due to the increased variability in requests from the mixed dataset, the problem of load imbalance, redundancy, and fragmentation is more severe. Similar to Section~\ref{sec:evaluation:e2e}, SGLang-MoonCake-6P2D and SGLang-MoonCake-2P6D struggle to handle multi-turn interactions, while SGLang-Router's attempt to improve hit rates leads to load imbalance and interference between the two phases. Ultimately, they fail to achieve 90\% SLO attainment. Additionally, under the same input token latency, \sysname's throughput is 5.47$\times$, 28.49$\times$, and 5.46$\times$ higher than that of SGLang-MoonCake-6P2D, SGLang-MoonCake-2P6D, and SGLang-Router, respectively. Furthermore, because only 25\% of prefix cache slots in SGLang-MoonCake-6P2D is available to its decoding phase, it has to frequently trigger recomputation, leading to a substantial output token latency.

%% file: sections/related.tex
\section{Related Work}
\label{sec:related}

\parabf{LLM serving.}
Modern LLM serving systems combine many techniques for better performance.
Continuous batching~\cite{yu2022orca} is proposed to batch different requests. To reduce interference between two phases,
chunked prefill~\cite{agrawal2024taming, kamath2025pod, holmes2024deepspeed} is proposed to split the long context into smaller chunks, while PD disaggregation~\cite{zhong2024distserve, patel2024splitwise, hu2024inference} is proposed to disaggregate two phases into different instances.
Elastic sequence parallelism~\cite{wu2024loongserve} is proposed to handle variable context lengths and frequent ratio changes between two phases by setting DoP elastically.
Many recent works also explore dynamic PD instance reconfiguration~\cite{wang2025prefill, qiao2024conserve, ruan2025dynaserve, jin2024system, du2025ecoserve, feng2025windserve, zhao2024blendserve} and multiplexing partial computation between two phases~\cite{liang2025injecting, shi2025nexus, hong2025semi, lin2025bullet, cui2025optimizing} to accommodate dynamic workloads.
These works focus on computation optimizations, while \sysname focuses on prefix cache pooling and provides declarative interfaces to support them.
AF disaggregation~\cite{xiao2025xdeepserve, chen2024efficient, he2024fastdecode, zhu2025megascale, stepfun2025step3largeaffordablemodelsystem} is proposed to disaggregate attention and FFN. Much research focuses on expert parallelism~\cite{zhao2025insights, he2021fastmoe, he2022fastermoe, lina, smartmoe} and sequence parallelism~\cite{shyam2024tree, agrawal2024medha, bhatia2025helix, yang2024context, brandon2023striped, liu2023ring} to efficiently parallelize the computation of FFN and attention modules across multiple devices. They are orthogonal to \sysname.

\parabf{KV cache optimizations.}  
Reusing KV caches is critical to accelerate LLM serving.  
PagedAttention~\cite{vllm} manages the KV cache as paged memory to reduce fragmentation.
SGLang~\cite{sglang} manages prefix cache as radix tree to enable efficient prefix sharing in an instance.
Many prefix caching systems, such as Mooncake~\cite{mooncake}, LMCache~\cite{lmcache}, Aibrix~\cite{aibrix}, Preble~\cite{preble}, and MemServe~\cite{memserve}, explore cache reuse across instances, while suffering from load imbalance, redundancy, and fragmentation as discussed in Section~\ref{sec:background}. Infinite-LLM~\cite{infinitellm} borrows the KV cache slots from other instances when local memory is insufficient, but it does not support prefix sharing between requests, leading to redundancy, uses the strict locality-based policy, leading to load imbalance, tightly couples the cache-free computation scheduling, restricting the elasticity, and cannot concurrently handle prefix attention and cache-free computation in an instance.
Many operators are developed to optimize prefix attention~\cite{flashinfer, ye2024chunkattention, song2024tackling}, and the cache policy is explored to optimize hit rate~\cite{wangkvcache}. They are orthogonal to \sysname. Many works also explore hierarchical KV cache management~\cite{yu2025stateful, gao2024cost, pan2025kvflow, xu2024pie, zuo2025serving, jiang2024neo}.
Because GPU memory often has higher bandwidth than host memory, and cache typically needs to be swapped back to GPU memory when used, \sysname can be used together with these solutions to further improve the performance.

%% file: sections/conclusion.tex
\section{Conclusion}
\label{sec:conclusion}

In this paper, we present \sysname, a unified segment-level prefix cache pool for elastic long-context LLM serving. By designing a declarative interface to expose query tensors and prefix attention, \sysname can manages prefix cache at the segment level to reduce fragmentation and enable fine-grained cache management. Based on it, we propose a heavy-hitter-aware segment-level load balancing algorithm to achieve better load balance and deduplication, and propose the bipartite matching-based dispatching algorithm to minimize the communication volume of query tensors and new prefix caches. Finally, \sysname enables stateless elastic scheduling to accommodate dynamic workloads. The evaluation results on real-world workloads demonstrate that \sysname significantly improves throughput and hit rate compared to existing cache-aware routing and cache-centric PD-disaggregation solutions.